\documentclass[pre,amsmath]{revtex4}
\usepackage{graphicx}
\usepackage{color}
\usepackage{bm}
\usepackage{epsfig}
\usepackage{latexsym}
\usepackage{pifont}
\usepackage{float}
\usepackage[utf8]{inputenc}
\graphicspath{ {figure/} }
\usepackage{siunitx}

\begin{document}

\newcommand{\be}{\begin{equation}}
\newcommand{\ee}{\end{equation}}
\newcommand{\bea}{\begin{eqnarray}}
\newcommand{\eea}{\end{eqnarray}}
\newcommand{\nn}{\nonumber}

\today

\title{Actin filaments pushing against a barrier: comparison between two force generation mechanisms}
\author{Raj Kumar Sadhu and Sakuntala Chatterjee}
\affiliation{Department of Theoretical Sciences, S. N. Bose National Centre for Basic Sciences, Block  JD, Sector  III, Salt Lake, Kolkata  700106, India. }

\begin{abstract}
To theoretically understand force generation properties of actin filaments, many models consider growing filaments pushing against a movable obstacle or barrier. In order to grow, the filaments need space and hence it is necessary to move the barrier. Two different mechanisms for this growth are widely considered in literature. In one class of models (type $A$), the filaments can directly push the barrier and move it, thereby performing some work in the process. In another type of models (type $B$), the filaments wait till thermal fluctuations of the barrier position create enough space between the filament tip and the barrier, and then they grow by inserting one monomer in that gap. The difference between these two types of growth seems microscopic and rather a matter of modelling details. However, we find that this difference has important effect on many qualitative features of the models. In particular, how the relative time-scale between the barrier dynamics and filament dynamics influences the force generation properties, are significantly different for type $A$ and $B$ models.  We illustrate these differences for three types of barrier: a rigid wall-like barrier, an elastic barrier and a barrier with Kardar-Parisi-Zhang dynamics. Our numerical simulations match well with our analytical calculations. Our study highlights the importance of taking the details of filament-barrier interaction into account while modelling force generation properties of actin filaments. 
\end{abstract}
\maketitle

\section{Introduction}
Cell motility has immense importance in many biological processes such as morphogenesis, wound repair, cancer invasion etc. \cite{review1, review2, review3, review4, review5}. Actin filaments play a crucial role in cell movement. The polymerization of actin filaments near the cell membrane form protrusions in the membrane, which is an important step for the cell movement. In this process, the actin filaments generate significant amount of force. This polymerization force is measured in many \textit{in vitro} cases by allowing these filaments to grow against some barrier and applying an opposing external force on that barrier \cite{marcy2004, baudry2011,theriot2005, mogilner2006,zimm, theriot2007}. The velocity of the barrier is expected to decrease with the external force and this force-velocity curve, is an important characteristic of the force generation mechanism. 

There have been intense research activities in the last few years to understand the force generation properties of actin filaments \cite{carlsson2014, baumgaertner2010, perilli2018,manoj2018, Sadhu2016, kirone, krawczyk2011, ddas2014, hansda2014, Sadhu2018, kolomeisky2015}. In many different experiments and  theoretical models, it has been investigated how polymerization of the filaments generate force that results in movement of the barrier. In our recent theoretical works, we have looked into how the force generation properties get affected by the (thermal) shape fluctuations of the barrier \cite{Sadhu2016,Sadhu2018}. We have shown that the shape fluctuations of the barrier have an important role to play and they affect the force-velocity curve significantly. In particular, the time-scale over which the shape fluctuations of the barrier takes place, can be smaller or larger compared to the time-scale of filament (de)polymerization process and this decides the qualitative shape of the force-velocity curve. In this paper, we show that not only this time-scale, but the microscopic modelling details of the interaction between the filament and the barrier can also be important. We illustrate this issue further in the following paragraph.

Consider a filament that polymerizes and pushes against a barrier. Now, there can be two different ways to model the interaction between the barrier and the filament. In one class of models, it is assumed that the thermal motion of the barrier creates a gap between the filament tip and the barrier position and a monomer gets inserted in that gap and the length of the filament thus grows. In \cite{peskin1993, carlsson2014, baumgaertner2010, perilli2018, manoj2018} such a mechanism was used. In a different set of models \cite{Sadhu2016, kirone, krawczyk2011, ddas2014, hansda2014, Sadhu2018, kolomeisky2015, mogilner1996}, it has been assumed that even when there is no gap between the filament tip and the barrier, the filaments can actively push the barrier away and grow. While the various models in \cite{peskin1993, carlsson2014, baumgaertner2010, perilli2018, manoj2018,Sadhu2016, kirone, krawczyk2011, ddas2014, hansda2014, Sadhu2018, kolomeisky2015, mogilner1996}   vary in many aspects, the interaction between the growing end of the filament and the barrier can be roughly classified in the above two types. Although the difference between the two mechanisms seems to be microscopic and rather a matter of modelling details, we find that the force generation properties are significantly different in the two cases. Interestingly, this difference persists even in absence of shape fluctuations of the barrier, when the barrier is modelled like a rigid wall. 

In this paper, we carry out a detailed quantitative study to highlight the difference between the above two mechanisms. For ease of nomenclature, we denote the two cases by type $A$ and type $B$, where type $A$ stands for those models where filaments can actively push the barrier and grow, and type $B$ represents those models where the filaments can grow only if there is sufficient gap between the filament tip and the barrier to insert one monomer. We demonstrate the difference between type $A$ and $B$ models for three different barrier dynamics. We have chosen particularly simple  models for the barrier dynamics and the fact that even for such simple systems, type $A$ and $B$ yield qualitatively different results, is rather remarkable. We consider (a) a rigid barrier model which has been widely studied earlier, including in \cite{carlsson2014, peskin1993, perilli2018, manoj2018, kirone, ddas2014, hansda2014}, (b) an elastic barrier model, which was studied in \cite{AVolmer1998, lipowsky94, baumgaertner2010, Sadhu2018,baumgaertner2012}, and (c) a Kardar-Parisi-Zhang (KPZ) barrier model, which was studied by us earlier \cite{Sadhu2016}. In all these three cases, the barrier dynamics is such that there is a competition between the barrier movement and filament growth. For the rigid barrier and KPZ barrier, an external force is applied on the barrier which tends to move the barrier in the direction opposite to that of filament polymerization. In the elastic barrier model, on the other hand, the barrier tries to minimize its elastic energy by remaining in a flat shape, while the filaments polymerize by creating protrusions on the barrier. For these three types of barrier dynamics, we compare the behavior of certain quantities for above two different growth conditions {\sl viz.} type $A$ and $B$. Some of the quantities we measure here for the first time, and some were known from earlier studies. But their comparison highlights the importance of difference between the two growth conditions. This comparison is important, since it was not known or appreciated before that a seemingly small difference like this can result in significantly different qualitative behavior.

For a rigid barrier with a type $A$ interaction with the filaments, we find the shape of force-velocity curve is always concave when there are multiple filaments present in the system. On the other hand, for a single filament, the force velocity curve is convex (concave) when the thermal fluctuations of the rigid barrier position are much slower (faster) than the filament dynamics. For type $B$ interaction, the force-velocity curve for multiple filaments shows convex (concave) shape for slower (faster) barrier dynamics, but for a single filament the shape does not depend on the time-scale.  We find that as the barrier dynamics becomes faster, $F_s$ decreases for type $A$, but remains constant for type $B$. Similar behavior for stall force was also found for a KPZ barrier. As the number of filaments in the system becomes larger, $F_s$ also increases. Our analytical calculations show that for type $B$, stall force scales linearly with the number of filaments, while for type $A$, there is a logarithmic correction to linear scaling. 

For an elastic barrier, as the membrane tension $\mu$ is varied, the velocity changes non-monotonically and shows a peak at $\mu = \mu^\ast$ \cite{Sadhu2018}. For  $\mu > \mu^\ast$, velocity decreases with $\mu$ and the system has a steady state. But for $\mu < \mu^\ast$ velocity increases with $\mu$ and our earlier studies in \cite{Sadhu2018} show that the system does not have a steady state in this regime. When the elastic barrier interacts with the filaments via type $A$ mechanism, $\mu^*$ increases as the barrier dynamics becomes slower. For very slow barrier dynamics, type $A$ model yields a monotonically decreasing convex $\mu - V$ curve. For type $B$, on the other hand, $\mu^*$ remains same for a wide range of the barrier movement time-scale and the shape of the $\mu - V$ curve remains non-monotonic. However, in the limit when the barrier motion becomes very fast, both the models show concave $\mu - V$ curves.

This paper is organized as follows. In Sec. \ref{sec:rigid} we present our results for the rigid barrier model, in Sec. \ref{sec:elastic} we discuss elastic barrier model, in Sec. \ref{sec:kpz} we discuss KPZ barrier model and finally present our conclusions in Sec. \ref{sec:sum}.

\section{Rigid barrier model} \label{sec:rigid}
\begin{figure}[h!]
\includegraphics[scale=1.0]{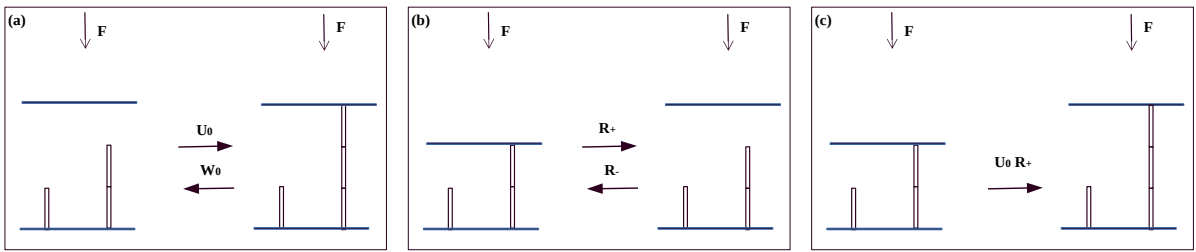}
\caption{Schematic representation of rigid barrier model. \textbf{(a):} A free filament polymerizes (depolymerizes) with rate $U_0$ ($W_0$). \textbf{(b):} The barrier height undergoes thermal fluctuations. In presence of the external force $F$, the forward process which increases the height is energetically costly and happens with rate $R_+$ while the reverse process that decreases the height is energetically favourable and happens with rate $R_-$. Using local detailed balance condition, we choose $R_+ = e^{-\beta F d} $ and $R_-=1$. \textbf{(c):} A bound filament pushes the barrier by an amount $d$ that costs energy and occurs with rate $U_0 R_+$. This movement is allowed only for type $A$.}
\label{model_rigid}
\end{figure}

In this model, the barrier is described as a one dimensional rigid and movable wall. Under the action of an external force $F$, the wall tends to move downward while thermal fluctuations can also move the barrier in the upward direction (see Fig.  \ref{model_rigid}b). The filaments are modelled as rigid rod-like objects, which are composed of rod-shaped monomers. For simplicity, we throughout assume the length of the monomer is same as $d$, the step size of barrier movement. A filament whose tip is in contact with the barrier is called a bound filament while in the absence of any such contact, it is called a free filament. Different types of filament movements are shown in Fig.  \ref{model_rigid}a and  \ref{model_rigid}c. For type $B$ model, the movement shown in Fig.  \ref{model_rigid}c is not allowed, while for type $A$ model, both the filament movements are allowed. We do not consider any dynamics at the base of the filaments and just assume that these bases are fixed at some hard wall. In real systems, however, the `hard wall' may be a densely connected actin network having a gel-like structure, but we do not include such descriptions in our simple model.

We perform simulations using kinetic Monte Carlo (MC) technique here. For a system having $N$ filaments, each Monte Carlo time-step consists of $N$ filament updates and ${\cal S}$ independent barrier updates, where ${\cal S}$ is the relative time-scale between the filament and the barrier dynamics. Starting from an initial condition of all filaments of length $d$ and the barrier in contact with all filament tips, the system undergoes time-evolution and after a large number of Monte Carlo steps, when the system reaches steady state, we perform our measurements. The various simulation parameters are given in Table \ref{table}. 

\subsection{Force-velocity curves for different values of ${\cal S}$} \label{sec:fv}
\begin{figure}[h!]
\includegraphics[scale=1.4]{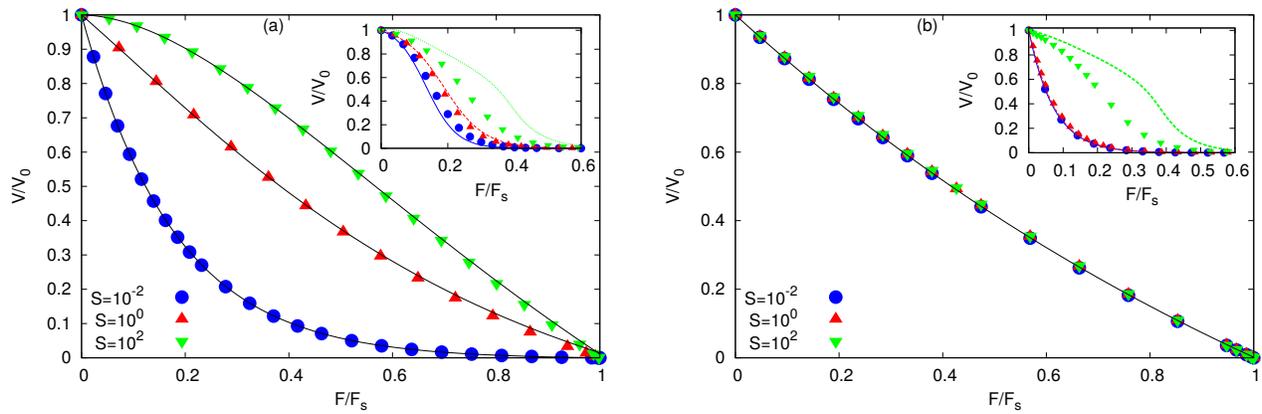}
\caption{$F-V$ curves for different values of ${\cal S}$ for rigid barrier. We plot the scaled velocity $V/V_0$ of the barrier as a function of the scaled force $F/F_s$, where $V_0$ is the velocity at $F=0$ and $F_s$ is the stall force. The main plot is for single filament while the inset is for multiple filaments. \textbf{(a)} Type $A$:  We note that  the force-velocity curve changes its nature with ${\cal S}$. For ${\cal S} <<1$, the curve is convex which changes to concave for ${\cal S} >> 1$. Inset: The force velocity curves is concave for any value of ${\cal S}$. \textbf{(b)} Type $B$: In this case, unlike type $A$, the shape of the force velocity curve is independent of $\cal S$. Inset: Force velocity curve show convex nature for small ${\cal S}$ while it changes to concave for large ${\cal S}$. Our analytical predictions are shown by continuous lines, that match well with numerics, but deviates for multiple filaments with large ${\cal S}$. For the insets, we use $N=20$. Simulation parameters are as in Table \ref{table}.}
\label{rigid_f_v_curve} \end{figure}

We measure the velocity of the barrier as a function of external force $F$ and plot in Fig. \ref{rigid_f_v_curve}. The main plot is for single filament while the inset is for multiple filaments. For single filament with type $A$ interaction, the force-velocity curve is convex for ${\cal S} \ll 1$ and becomes concave for large $\cal S$ \cite{carlsson2014}. However, for type $B$ model, our data in Fig.  \ref{rigid_f_v_curve}b show that the force-velocity curve does not change shape with $\cal S$. We explain this observation below by analytically calculating $V$ as a function of $F$.

For type $A$ systems with single filament, the expression for velocity can be written as $V (F)=d\{p_0 U_0 e^{-\beta F d} + {\cal S}e^{-\beta F d} - {\cal S} (1-p_0)\}$. Here, $p_0$ is the contact probability of the filament tip with the barrier. The first term corresponds to the bound filament polymerization (Fig.  \ref{model_rigid}c); the second and the third terms are for upward and downward thermal fluctuations of the barrier respectively (Fig.  \ref{model_rigid}b). The downward movement occurs only when the filament is not in contact with the barrier, thereby multiplied by a factor $(1-p_0)$. The value of $p_0$ can be calculated by solving the master equations for $p_i$ in the steady state, where $p_i$ is the probability that there is a gap of magnitude $i$ in between the filament tip and the barrier. We thus obtain, $p_0=\frac{U_0-W_0+{\cal S} (1-e^{-\beta Fd})}{U_0+{\cal S}}$. Using the expression for $p_0$ we get 
\be
V(F)=d\bigg[ \frac{ (U_0e^{-\beta F d} + {\cal S}) \{U_0-W_0+{\cal S} (1-e^{-\beta F d})\}}{(U_0+{\cal S})} - {\cal S}(1-e^{-\beta F d})\bigg]. 
\label{eq:rigid_ebr_v}
\ee
Our numerical data in Fig. \ref{rigid_f_v_curve}a show good agreement with Eq. \ref{eq:rigid_ebr_v}.

For type $B$, velocity has the form $V(F)=d\{{\cal S}e^{-\beta F d} - {\cal S} (1-p_0)\} $, which after putting the value of $p_0$ becomes,
\be 
V(F)=\frac {d\{(U_0-W_0) - U_0(1-e^{-\beta F d})\}}{1+ \frac{U_0}{{\cal S}}}
\label{eq:rigid_br_v}
\ee
and if we scale the velocity by $V_0$, which is the velocity at $F=0$, we have, $\frac{V(F)}{V_0}=\frac{U_0 e^{-\beta F d}-W_0}{U_0-W_0}$ which is independent of ${\cal S}$. Thus, the shape of the force-velocity curve does not depend on $\cal S$ in this case (Fig. \ref{rigid_f_v_curve}b). 

Stall force $F_s$ is defined as the external force, which exactly balances the force generated by the filaments. At $F=F_s$, the barrier velocity therefore becomes zero, and hence can be obtained by simply putting $V(F)=0$ and solving for $F$, which gives

\be
F_s=-\frac{1}{\beta d} ln \bigg [ \bigg( \frac{U_0-W_0+2{\cal S}}{2{\cal S}}\bigg ) \bigg(1-\sqrt{1-\frac{4U_0W_0{\cal S}^2}{U_0^2(U_0-W_0+2{\cal S})^2}} \bigg) \bigg] 
\label{eq:single_fs}
\ee
for type $A$, which decreases with $\cal S$. For type $B$ model, we have $F_s=\frac{1}{\beta d} ln \frac{U_0}{W_0}$, which is independent of $\cal S$. 

Our calculations for single filament can be easily generalized for multiple filaments and in Fig. \ref{rigid_f_v_curve} insets, we show the results using continuous lines, which show good agreement with our numerics. The discrepancy at large $\cal S$ can be attributed to the fact that in our analytical calculation, we have used the assumption that the filaments are independent. This is not strictly true, since the barrier induces an effective coupling between the filaments and for large ${\cal S}$ the effect of this coupling can not be neglected any more. 

Calculation of stall force for multiple filament case yields, 
\be
F_s=\frac{1}{\beta d} \Big \{N \ln \Big ({\frac{U_0}{W_0}} \Big ) + \ln \Big (1+\frac{N(U_0-W_0)}{{\cal S}}\Big ) \Big \} 
\label{eqn:fs_na}
\ee
for type $A$ model. Polymerization of one bound filament causes movement of the barrier, and all other bound filaments are immediately free. This induces an effective coupling between the filaments and $F_s$ does not scale with $N$ any more. For type $B$ model, however, this effect is absent and we find $F_s=\frac{N}{\beta d} \ln ({\frac{U_0}{W_0}} )$. In  \cite{ddas2014, carlsson2008}, a similar absence of linear scaling between $F_s$ and $N$ was reported and was explained by using effects like ATP hydrolysis of actin monomers. 
\section{Elastic barrier model} \label{sec:elastic}

In this section, we discuss the difference between type $A$ and $B$ dynamics for an elastic barrier model. The barrier or membrane is modelled as a one dimensional lattice of length $L$ and lattice constant $d$. At each site $i$ of the lattice a height $h_i$ is assigned. The membrane prefers to stay flat, when height of all sites are the same. Any height gradient costs elastic energy which can be calculated as \cite{lipowsky94, AVolmer1998, baumgaertner2010, Sadhu2018}
\be
{\cal H}_1=\mu \sum_{i=1}^{L}|h_i-h_{i+1}|
\label{eq:hamiltonian} 
\ee
where $\mu$ is the membrane tension. The sum in Eq. \ref{eq:hamiltonian} is related to the total contour length $\cal C$ of the particular height configuration of the membrane, such that ${\cal C} =\sum_{i=1}^{L}|h_i-h_{i+1}| + Ld$. This model was used in \cite{lipowsky94} to study the behavior of lipid membrane protrusions created due to thermal fluctuations. In \cite{baumgaertner2010} a similar model was used to study a membrane driven by an advancing actin mesh. 

\begin{figure}[h!]
\includegraphics[scale=1.3]{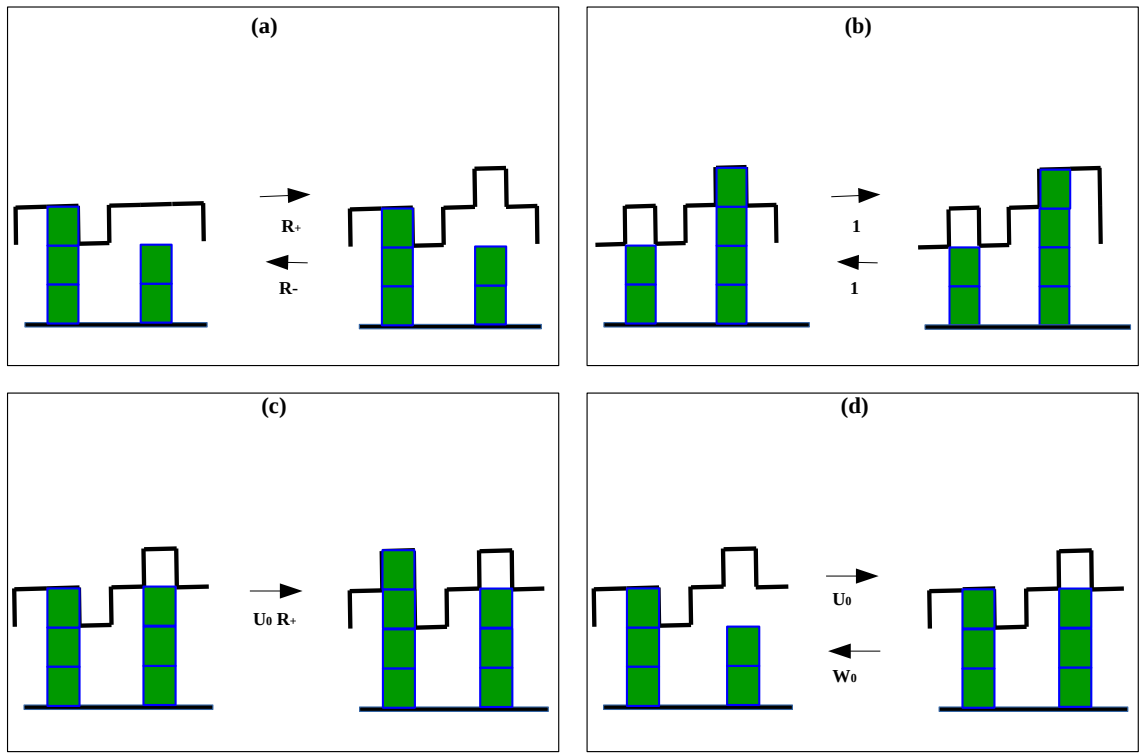}
\caption{Schematic representation of elastic barrier model. The square blocks show actin monomers, which join together to form rod-like filaments. The thick solid line represents the shape of the elastic membrane. \textbf{(a):} A bulk site of the membrane thermally fluctuates and changes its height by an amount $d$, which in turn changes the membrane contour length by $2d$. The forward process increases the energy and occur with rate $R_+$ while the reverse process decreases the energy and happens with rate $R_-$. \textbf{(b):} A bulk site of the membrane changes its height by an amount $d$ but the energy remains same and thus the movement happens with rate unity. \textbf{(c):} A bound filament pushes the binding site by an amount $d$ that costs energy and occurs with rate $U_0 R_+$. This movement is allowed only for type $A$. \textbf{(d):} A free filament polymerizes (depolymerizes) with rate $U_0$ ($W_0$).}
\label{model}
\end{figure}

In our model, we assume that the thermal fluctuations of the membrane height follow local detailed balance with the above Hamiltonian 
\be
\frac{R_+}{R_-}=e^{-\beta  \Delta E} \label{eq:ldb}
\ee
where $R_+$ ($R_-$) is the rate of those processes that increases (decreases) the energy by an amount $\Delta E$ (see Fig. \ref{model}a). In our lattice model, we assume that as a result of these fluctuations, the local height can increase or decrease by a discrete amount $d$ and the energy cost $\triangle E$ can take the values $2\mu d$ and $0$. 
In our simulation, we choose $R_+=e^{-\beta \mu d}$ and $R_-=e^{\beta \mu d}$ when $\triangle E = 2 \mu d$, and for $\triangle E=0$, we choose $R_+=R_-=1$ (for example, Fig. \ref{model}b). Note that any height fluctuation that brings a binding site at a lower height than the filament tip is forbidden. Everywhere else in the membrane, the height fluctuations will occur in accordance with Eq. \ref{eq:ldb}. 

The filaments are modelled as rigid rod-like polymers, composed of monomers of length $d$. The filament dynamics is identical to what was described in Sec. \ref{sec:rigid}. As before, there are free filaments and bound filaments. The bound filaments touch the membrane at the `binding sites'. For type $A$ model, polymerization of a bound filament 
increases the height of the binding site by an amount $d$ and hence an energy cost $\Delta E$ is involved (see Fig. \ref{model}c). Note that the change in energy can be positive or negative, or even zero, depending on the local height configuration around the binding site. For a positive (negative) energy cost, the bound filament polymerization rate is taken as $U_0 R_+$ ($U_0 R_-$), while for no energy cost, the rate is $U_0$.

The simulation technique is same as described for the rigid barrier case. For a system of $N$ filaments, in each Monte Carlo step, we attempt $N$ filament updates and ${\cal S}$ independent membrane updates. By changing $\cal S$, we can tune the relative time-scale between the filament dynamics and the membrane dynamics. Since the membrane consists of $L$ sites in the present case, while updating the membrane, we choose any site randomly and then follow the update rules as described. We present our results for the single filament case here and even for multiple filaments (with small or moderate density) we find similar behavior (data not shown here).

In an earlier study \cite{Sadhu2018} we had considered an elastic barrier with type $A$ interaction, and we have borrowed some data from this study here for the sake of comparison. In \cite{baumgaertner2010} an elastic barrier with type $B$ interaction was studied but only in the limit of $N=L$, i.e. when there is a filament pushing against all barrier sites. As we argue below, this limit ensures that the system always reaches a steady state \cite{Sadhu2018}. Here, we consider the case $N < L$ and find absence of steady state for small $\mu$.

\subsection{Variation of $V_0$ with ${\cal S}/L$}
\begin{figure}[h!]
\includegraphics[scale=0.7]{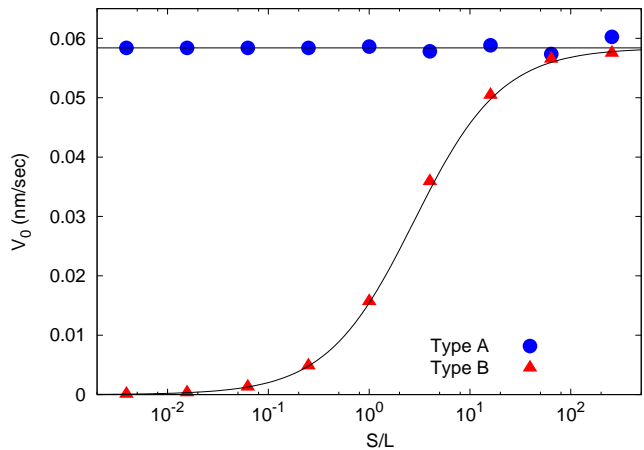}
\caption{Velocity at $\mu=0$ as a function of ${\cal S}/L$ for elastic barrier. For type $A$, we note that the velocity for $\mu=0$ remains constant with ${\cal S}/L$. The value is $V_0=d (U_0-W_0)/L$. For type $B$, unlike type $A$, velocity increases with ${\cal S}/L$ and saturates to a value $d(U_0-W_0)/L$, same as the velocity for type $A$. We use $L=64$ for both the cases. Other simulation parameters are as in Table \ref{table}.}
\label{s_vs_v_0} \end{figure}

We measure the velocity of the barrier when the elastic tension $\mu=0$ and denote it by $V_0$. For $\mu=0$, there is no energy cost for creating local protrusions or height gradient on the membrane. Apart from the binding sites, all $(L-1)$ membrane sites undergo equilibrium thermal fluctuations in this case and the time-averaged local velocity of the membrane is zero at all these sites. Only the binding site feels the presence of the growing filament: for type $A$ interaction, the filament pushes at the binding site and gives it a non-zero velocity while for type $B$ interaction, the filament blocks downward movement of the binding site and thus imparts a velocity to the membrane. The expression for $V_0$ for type $A$ is thus, $V_0=\frac{d}{L}\{U_0p_0+{\cal S}-{\cal S}(1-p_0)\}$, where $p_0=\frac{U_0-W_0}{U_0+{\cal S}}$ is the contact probability which is calculated in a similar way as for the rigid barrier case. Using the value of $p_0$ we have, for type $A$ model, $V_0 = d(U_0-W_0)/L $, which does not depend on $\cal S$. For type $B$ model, we have $V_0 = \frac{d}{L} \frac{(U_0-W_0)}{1+\frac{U_0}{({\cal S}/L)}}$, which has an explicit dependence on $\cal S$. In Fig. \ref{s_vs_v_0} we show the comparison with simulation data. 


\subsection{$\mu-V$ curves for different values of ${\cal S}/L$}

Different values of local velocities at bulk site and binding sites of the membrane at $\mu =0$ implies that the system does not have a steady state in this case. Even for non-zero but small values of $\mu$ this difference persists and the system fails to reach a steady state. Finally, when $\mu$ is larger than a certain threshold value $\mu^\ast$, the local velocity becomes same everywhere on the membrane and the system reaches a steady state. In an earlier work \cite{Sadhu2018} we have presented a detailed discussion on this interesting phenomenon. Here, we briefly describe the main points.

First let us consider ${\cal S} / L \sim 1$. As the filaments grow and push the membrane, there is a competition between the elastic force and the polymerization force. For small $\mu$, the elastic force loses out and the polymerization force wins. In this case, the binding site (which is mainly driven by the polymerization force) moves upward with a larger velocity $V_{bind}$ and the bulk sites (which are purely driven by the elastic force) move with smaller velocity $V_{bulk}$. The bulk sites move upward, trying to catch up with the binding site and thereby decreasing the contour length of the membrane. As $\mu$ increases, the elastic force becomes stronger and the velocity of the bulk sites increases, while the velocity of the binding site decreases due to the increase in the elastic energy cost. The average velocity of the membrane, which can roughly be written as $VL= (L-1) V_{bulk} + V_{bind}$, also increases with $\mu$ in this regime. Finally, for $\mu = \mu^\ast$ the two velocities become equal, $V_{bind} = V_{bulk}$ and for $\mu$ larger than this threshold value, the system has a steady state when the whole membrane moves up with the same velocity. The membrane velocity decreases with $\mu$ in this regime. From the above discussion,  it is also clear that $V_{bind}$ falls monotonically with $\mu$ while $V_{bulk}$ shows a non-monotonicity, with a peak at $\mu=\mu^*$. For a more elaborate explanation of this mechanism see \cite{Sadhu2018}.

\begin{figure}[h!]
\includegraphics[scale=1.4]{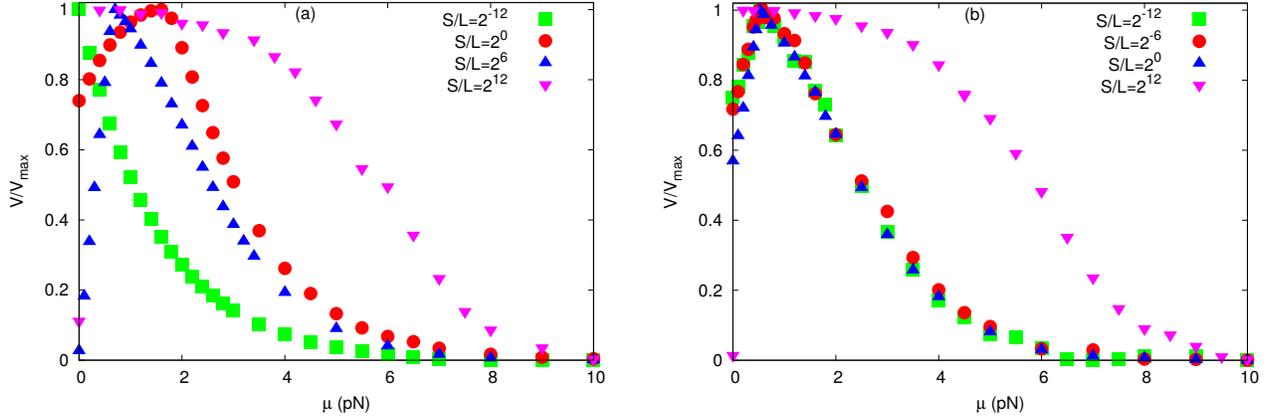}
\caption{$\mu-V$ curve for different values of ${\cal S}/L$ for elastic barrier. We have scaled $V$ by maximum velocity $V_{max}$ to compare them in the same scale. \textbf{(a)} Type $A$: For ${\cal S}/L \sim 1$, the $\mu-V$ curve is non-monotonic with a peak at $\mu=\mu^*$. As ${\cal S}/L$ increases, the peak shifts towards to smaller value of $\mu$ and finally the curve becomes concave for ${\cal S}/L >> 1$.  For ${\cal S}/L <<1$, the curve is convex. \textbf{(b)} Type $B$: In this case, unlike the previous case, we do not have any convex curve even for ${\cal S}/L <<1$.  For all the above plots, we have used  $L=64$. Note that for ${\cal S}/L >>1$, both the cases show a concave $\mu-V$ curve. Other simulation parameters are as in Table \ref{table}.}
\label{mu_v_curve} \end{figure}

In Fig. \ref{mu_v_curve} we plot $\mu - V$ curve for different values of ${\cal S}/L$. The $\mu-V$ data for type $A$ interaction were presented earlier in 
 \cite{Sadhu2018}. For both type $A$ and $B$ dynamics, the curve shows a peak for ${\cal S} / L =1$ and becomes monotonic concave curve for ${\cal S} / L \gg 1$. However, as ${\cal S}/L$ becomes smaller than unity, the curve becomes convex for type $A$ and for type $B$ the curve continues to show a peak and the peak position does not depend on ${\cal S}/L$ in this regime. So the main qualitative difference in this case is obtained for small values of ${\cal S}/L$ when the membrane dynamics is much slower than the filament dynamics. Since for very large $\cal S$, the difference between type $A$ and $B$ ceases to exist, we find concave $\mu -V$ curves for both models in this limit.

\begin{figure}[h!]
\includegraphics[scale=0.7]{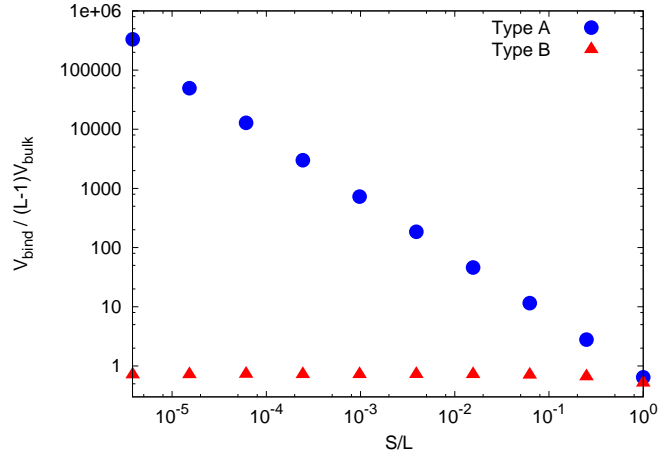}
\caption{Variation of $\frac{V_{bind}}{(L-1)V_{bulk}}$ with ${\cal S}/L$ for elastic barrier. For type $A$, we note that as ${\cal S}/L$ decreases, the value of  $\frac{V_{bind}}{(L-1)V_{bulk}}$ increases which means that the contribution of binding site to the total velocity increases in comparison to the bulk sites. Thus, the nature of $\mu-V$ curve changes to convex for ${\cal S}/L << 1$. Here we use $\mu=0.87 pN$ which is less than $\mu^\ast$. For type $B$, as ${\cal S}/L$ decreases, the value of  $\frac{V_{bind}}{(L-1)V_{bulk}}$ does not vary much. Thus, even for ${\cal S}/L<<1$, the terms $V_{bind}$ and $V_{bulk}$ remain comparable and the $\mu-V$ curve remains non-monotonic. Here we use $\mu=0.26 pN$. For both the models, we use $L=64$. Other simulation parameters are as in Table \ref{table}.}
\label{vb_by_v_bulk} \end{figure}

In order to explain this difference, we measure the relative contribution of $V_{bind}$ and $V_{bulk}$ towards the average membrane velocity. In Fig. \ref{vb_by_v_bulk} we plot $\frac{V_{bind}}{(L-1)V_{bulk}}$ at a fixed $\mu < \mu^*$ as a function of ${\cal S}/L$. When this ratio is close to $1$, that means both binding site and bulk sites have comparable contribution, but if the ratio is much larger than $1$,  bulk site contribution can be ignored. Our data show that for type $A$ model, the ratio increases as ${\cal S}/L$ decreases, meaning for small ${\cal S}/L$ binding site velocity dominates. Since, $V_{bind}$ falls monotonically with $\mu$ \cite{Sadhu2018}, the average velocity also falls monotonically. For type $B$, on the other hand, the ratio remains close to $1$ even when ${\cal S}/L <<1$ and due to non-monotonic variation of $V_{bulk}$ with $\mu$ the $\mu - V$ curve remains peaked in this case.

\subsection{Variation of $\mu^*$ with ${\cal S}/L$}
\begin{figure}[h!]
\includegraphics[scale=0.7]{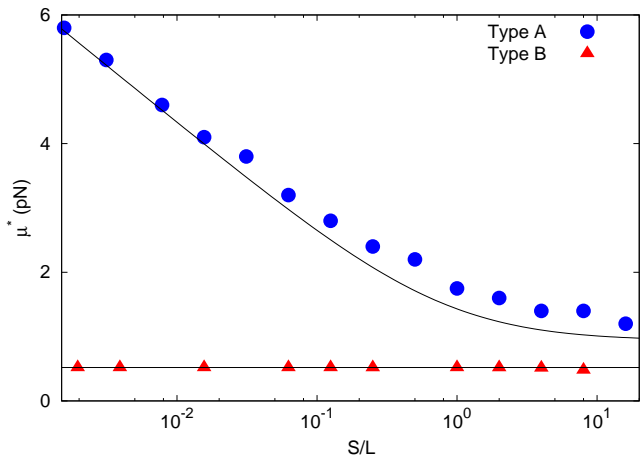}
\caption{Variation of $\mu^*$ with ${\cal S}/L$ for elastic barrier. For type $A$, $\mu^*$ decreases continuously with ${\cal S}/L$. For type $B$, $\mu^*$ remains constant for a wide range of ${\cal S}/L$. Our analytical predictions matches well with the numerical results. Simulation parameters are as in Table \ref{table}.}
\label{mu_star} \end{figure}

The elasticity-velocity curve in Fig. \ref{mu_v_curve} show that the variation of $\mu^*$ as a function of ${\cal S}/L$ is different for types $A$ and $B$. We note that for type $B$, the peak position in the elasticity-velocity curve remains fixed over a wide range of ${\cal S}/L \lesssim 1$, but for type $A$, the peak shifts towards the smaller $\mu$ as ${\cal S}/L$ increases, which means $\mu^*$ decreases with ${\cal S}/L$. This behaviour can be explained by the force balance at the tip of filament. Note that for the rigid barrier case, the balance between polymerization force and externally applied force is reached at the point of stalling and we can therefore use the expression for $F_s$ calculated in Eq. \ref{eq:single_fs} for the polymerization force generated by the filaments, even in the case of an elastic barrier.  The only modification in this case is for type $A$ model, where there is an explicit $\cal S$ dependence of $F_s$ (see Eq. \ref{eq:single_fs}), one must replace 
$\cal S$ by ${\cal S}/L$ because unlike a rigid barrier, the whole elastic barrier does not move as a single unit and height fluctuations take place at each of $L$ sites on the barrier. This gives for type $A$
\be
\mu^*=-\frac{1}{2 \beta d} ln \bigg [ \bigg( \frac{U_0-W_0+2\frac{\cal S}{L}}{2\frac{\cal S}{L}}\bigg ) \bigg(1-\sqrt{1-\frac{4U_0W_0(\frac{\cal S}{L})^2}{U_0^2(U_0-W_0+2\frac{\cal S}{L})^2}} \bigg) \bigg]
\ee
and for type $B$
\be 
\mu^*=\frac{1}{2 \beta d} ln \Big (\frac{U_0}{W_0} \Big ).
\ee
Our data in Fig. \ref{mu_star} shows good agreement with this calculation. This plot shows  that for type $A$, as ${\cal S}/L$ is increased, $\mu^*$ decreases and for type $B$ it remains constant. The data for type $A$ appeared earlier in \cite{Sadhu2018}. Note that although the $\mu -V$ curve becomes monotonic for type $A$ model in the limit of small ${\cal S}/L$, one still has a finite $\mu ^\ast$.

In this section, we have assumed the energy of the barrier to be proportional to the sum of local height gradients (or the contour length) of the barrier (Eq. \ref{eq:hamiltonian}). In another related model, where the elastic energy of the barrier is proportional to square of the local height gradient \cite{Sadhu2018, baumgaertner2012}, we find only quantitative difference between type $A$ and $B$ dynamics (data not shown here), and qualitative behavior remains mostly similar. We also note that in the limit of $\mu >> 1$, both the models of elastic barrier actually behaves like a rigid barrier \cite{Sadhu2018}.

\section{KPZ barrier model} \label{sec:kpz}
\begin{figure}[h!]
\includegraphics[scale=1.7]{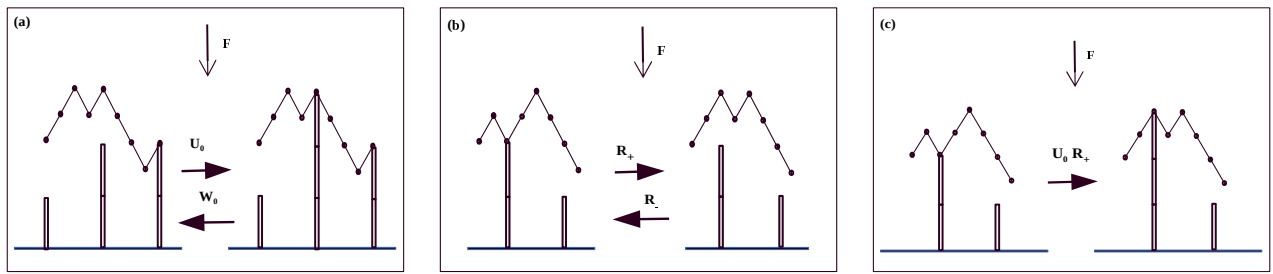}
\caption{Schematic representation of our model. (a): A free filament polymerizes and depolymerizes with rates $U_0$ and $W_0$, respectively. (b): A local valley can flip to a hill with rate $R_+$ and the reverse process occurs with rate $R_-$. (c): Polymerization of a bound filament that flips a valley to a hill happens with rate $U_0 R_+$. This particular movement is allowed for type $A$ only.}
\label{kpz_model} \end{figure}

In this section, we present the comparison between type $A$ and $B$ models for a KPZ barrier. The model consists of $N$ parallel filaments growing against a barrier with a fluctuating height profile whose dynamics is described by a one dimensional Kardar-Parisi-Zhang or KPZ interface \cite{kpz,Sadhu2016}. An external force $F$ is acting on the barrier in the opposite direction of the filament growth (see Fig. \ref{kpz_model}). In our lattice model, the discrete surface elements are represented as lattice bonds, which can have two possible orientations, $\pm \pi /4$. We denote these two cases by symbols $/$ and $\backslash$ and call them upslope and  
downslope bonds, respectively. The total number of such bonds is $L$. One $/$ followed by a $\backslash$ forms a local hill and in the reverse order $ \backslash /$ they form a local valley. The local height of the surface fluctuates due to transition between these hills and valleys. When a local hill (valley) at a given site flips to a valley (hill), the height of that particular site decreases (increases) by an amount  $d$, which is same as the length of an actin monomer.

In Fig. \ref{kpz_model} we show the update rules. The barrier shows thermal fluctuations, when local hills can flip to valleys and vice versa. When a valley flips to a hill, it has to work against the force $F/L$ (force per site). We assign a rate $R_+$ for this process. The reverse process, where a hill flips to a valley is energetically favourable and occurs with rate $R_-$. Using the local detail balance condition, we choose $R_+=e^{-\beta Fd/L}$ and $R_-=1$ (Fig. \ref{kpz_model}b). Note that the hill can only flip to a valley when no filament is in contact with that particular site. The filament dynamics is same as described in the earlier cases. A bound filament can polymerize for type $A$ model, only if it is in contact with a local valley which flips to a hill due to upward push by the filament. The rate of this process is $U_0 e^{-\beta Fd/L}$. For type $B$ model, bound filament polymerization is not allowed.

In an earlier work \cite{Sadhu2016}, we had studied KPZ-barrier with type $A$ interaction, but with a slightly different model, where bound filament polymerization could cause a global movement of the whole barrier, in addition to the local movement of a valley to hill that we consider here. For the model considered in \cite{Sadhu2016} a bound filament polymerization was possible even when there is no local valley present at the binding site. In this case a polymerization event would cause movement of the whole barrier (like a rigid object) and this happens with rate $U_0 e^{-\beta Fd}$. In the present work we are interested in comparison between type $A$ and $B$ mechanism and to ensure that the other differences between the models are minimal, we have not considered the global movement of the barrier here.

We assume periodic boundary condition for the surface and an equal number of upslope and downslope bonds. The simulation procedure is same as described in earlier cases. In one Monte Carlo step, we attempt to perform $N$ filament updates (polymerization or depolymerization) and ${\cal S}$ independent surface updates (hill-valley flipping). We let the system evolve for a long time, according to above dynamical rules and then perform the measurements in the steady state.

\begin{figure}[h!]
\includegraphics[scale=0.7]{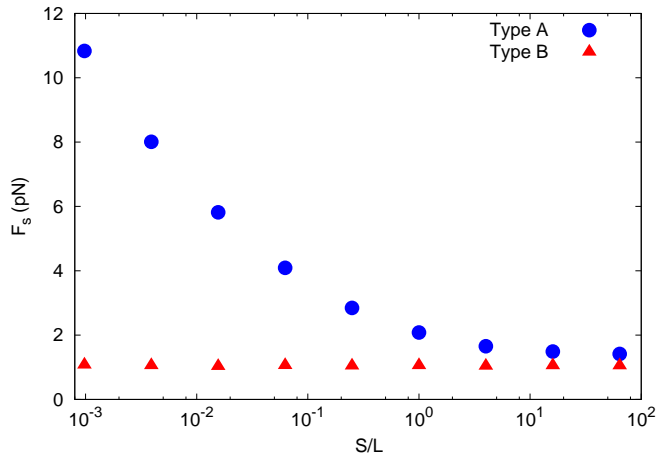}
\caption{$F_s$ as a function of ${\cal S}/L$. We note that for type $A$, the stall force decreases with ${\cal S}/L$. For type $B$, the stall force remains constant with ${\cal S}/L$ over a wide range of ${\cal S}/L$. For very large ${\cal S}/L$, $F_s$ for both the cases saturates to the same value. Here we use $L=64$, $N=1$.  Other simulation parameters are as in Table \ref{table}.}
\label{kpz_s_vs_fs} \end{figure}

We find that the barrier velocity is systematically higher for type $A$ compared to type $B$ model, as expected. However, in this case the main qualitative difference between the two models is observed only in the dependence of stall force $F_s$ on $\cal S$. In Fig. \ref{kpz_s_vs_fs} we show that for type $A$ stall force decreases with $\cal S$ and for type $B$ it stays constant. This is consistent with the result for rigid and elastic barrier, respectively. For KPZ barrier, many other quantities like $V_0$, force-velocity curve,  while quantitatively different for type $A$ and $B$ models, show similar qualitative behavior. The reason can be explained in the following way. Although bound filament polymerization is allowed for type $A$ model, this process happens only when there is a valley present at the binding site. The probability for such an occurrence is low. Most of the time the binding site sits on a local hill and when in contact with a filament, that hill cannot even flip to a valley. Thus even though type $A$ model allows bound filament polymerization, most of the time this polymerization does not take place and the mechanism then becomes effectively similar to type $B$.


\section{Conclusion}
\label{sec:sum}
In summary, we have studied the force generation by a set of parallel actin filaments growing against a barrier by considering two models for the filament-barrier interaction. In type $A$ model, the filaments can push the barrier and grow, while in type $B$ model, the filaments can grow only when thermal fluctuations of the barrier position creates a space for growth. It is clear from the definition of the two models that in identical conditions, the velocity of the barrier in type $A$ model cannot fall below that in type $B$ model. We also find important qualitative differences between the two model types. We demonstrate these differences for three barrier dynamics: rigid barrier, KPZ barrier and elastic barrier. In all three cases, we find that the main differences between type $A$ and $B$ models are manifested in the way various properties depend on the relative time-scale $\cal S$ between the filament and barrier dynamics. In the limit of very large $\cal S$, when the thermal fluctuations of barrier position are much faster than filament (de)polymerization, we find type $A$ and $B$ models become similar. In this case, since the contact probability $p_0$ becomes negligible, the filaments grow  freely most of the time without pushing the barrier even for type $A$ interaction. Thus, in this limit, the two models yield same results. However, for small and moderate $\cal S$ values, we find significant differences.

The competition between polymerization force of the filaments and the  opposing force exerted by the barrier shapes the underlying physics of the system. For a rigid barrier and a KPZ barrier, this opposing force is described as an external force that tends to move the barrier downward, while polymerization force coming from the filaments is pushing the barrier upward. The point of balance between these two forces is achieved at stalling, when the barrier velocity vanishes. For an elastic membrane, the competition between the two forces plays out in a different way. In this case, the elastic force of the barrier tries to maintain a flat shape of the membrane, while filament polymerization tends to create protrusions on the membrane. The point of balance in this case is $\mu^\ast$, when the binding site and bulk site velocities become equal. We show that for all three barriers, the point of balance depends on $\cal S$ for type $A$ model: the stall force $F_s$ (for rigid and KPZ barrier) and $\mu^\ast$ (for elastic barrier) decreases with $\cal S$ and saturates to a constant value for large $\cal S$. For type $B$ models, we find that $F_s$ and $\mu^\ast$ remains independent of $\cal S$. In other words, when the bound filament polymerization is not allowed, the polymerization force is insensitive to the time-scale of thermal fluctuations, but when the bound filaments can actively push the barrier and grow, the polymerization force decreases as the barrier dynamics becomes faster. On the contrary, when there is no opposing force offered by the barrier, i.e. $F=0$ or $\mu=0$, the two models show opposite behavior. For type $A$ models, in this limit there is no difference between free and bound filament polymerization rate, and the barrier velocity $V_0$ in this case does not depend on $\cal S$ for elastic barrier. However, for type $B$ models, since bound filaments cannot polymerize, polymerization process still has to wait when thermal fluctuations create a space for growth and hence $V_0$ for type $B$ models increases with $\cal S$. The same behavior of $V_0$ is also seen for a rigid barrier (data not shown).

Our study shows that while modelling force generation of actin filaments, it is important to consider what type of filament barrier interaction should be included in the model. The seemingly minor details like whether bound filament polymerization is allowed or not, may give rise to qualitatively different outcomes. Based on the results we have presented here, it is also possible to perform experimental measurements to decide which type of model would be suitable in a given situation. The relative time-scale between the barrier and filament dynamics may be varied by controlling the free filament polymerization rate which depends on the concentration of monomers in the medium. By directly measuring the variation of quantities like stall force, or force-velocity curve as a function of this time-scale, it should be possible to decide the suitability of a particular model type.

\section{Acknowledgements}
SC acknowledges financial support from the Science and Engineering Research Board, India (Grant No. EMR/2016/001663). The computational facility used in this work was provided through Thematic Unit of Excellence on Computational Materials Science, funded by Nanomission, Department of Science and Technology, India. 
\begin{table}
\begin{center}
 \begin{tabular}{|c | c| c|}
 \hline
 $U_0$ & Free filament polymerization rate & $2.784 s^{-1}$\\
 \hline
 $W_0$ & Filament depolymerization rate & $1.4 s^{-1}$ \\
 \hline
 $d$ & Size of an actin monomer & $2.7 nm$ \\
 \hline
 $T$ & Temperature & $300 K$ \\
 \hline
\end{tabular}
 \caption{Parameters used in our simulation. The values of $U_0$ and $W_0$ are taken from \cite{pollard} and the value of $d$ is taken from \cite{hansda2014}.}
 \label{table}
\end{center}
\end{table}


\begin{thebibliography}{99}
\bibitem{review1} J. Howard, Mechanics of motor proteins and the cytoskeleton, Sunderland, MA: Sinauer Associates (2001).

\bibitem{review2} T. D. Pollard and J. A. Cooper, Actin, a central player in cell shape and movement, Science {\bf 326}, 1208 (2009).

\bibitem{review3} L. Blanchoin, R. B. Paterski, C. Sykes and J. Plastino, Actin dynamics, architecture and mechanics in cell motility, Physiol Rev {\bf 94}, 235 (2014). 

\bibitem{review4} P. Friedl and D. Gilmour, Collective cell migration in
morphogenesis, regeneration and cancer, Nat. Rev. Mol. Cell Biol. {\bf 10}, 445 (2009).

\bibitem{review5} J. Plastino and C. Sykes, The actin slingshot, Curr. Opin. Cell Biol. {\bf 17}, 62 (2005)

\bibitem{marcy2004}  Y. Marcy, J. Prost, M. F. Carlier, and C. Sykes, Forces generated during actin-based propulsion: A direct measurement by micromanipulation, Proc. Natl. Acad. Sci. U.S.A. {\bf 101}, 5992 (2004).

\bibitem{baudry2011} C. Brangbour, O. du Roure , E. Helfer, D. Démoulin, A. Mazurier, M. Fermigier, M. F. Carlier, J. Bibette and J. Baudry, Force-velocity measurements of a few growing actin filaments, PLoS Biol. {\bf 9}, e1000613 (2011). 

\bibitem{theriot2005} S. H. Parekh, O. Chaudhuri, J. A. Theriot and D. A. Fletcher, Loading history determines the velocity of actin-network growth, Nat. Cell Biol. {\bf 7}, 1219 (2005).

\bibitem{mogilner2006}  M. Prass, K. Jacobson, A. Mogilner and M. Radmacher, Direct measurement of the lamellipodial protrusive force in a migrating cell, J. Cell. Biol. {\bf 174}, 767 (2006). 

\bibitem{zimm} J. Zimmermann, C. Brunner, M. Enculescu, M. Goegler, A. Ehrlicher, J. K{\"a}s  and M. Falcke, Actin filament elasticity and retrograde flow shape the force-velocity relation of motile cells, Biophys. J. {\bf 102}, 287 (2012).

\bibitem{theriot2007} M. J. Footer, J. W. J. Kerssemakers, J. A. Theriot and M. Dogterom, Direct measurement of force generation by actin filament polymerization using an optical trap,  Proc. Natl. Acad. Sci. U.S.A. {\bf 104}, 2181 (2007). 

\bibitem{carlsson2014} R. Wang and A. E. Carlsson, Load sharing in the growth of bundled biopolymers, New J. Phys {\bf 16}, 113047 (2014).

\bibitem{baumgaertner2010} S. L. Narasimhan and A. Baumgaertner, Dynamics of a driven surface, J. Chem. Phys. {\bf 133}, 034702 (2010).

\bibitem{perilli2018} A. Perilli, C. Piereoni, G. Ciccotti and J. P. Ryckaert, On the force-velocity relationship of a bundle of rigid bio-filaments, J. Chem. Phys \textbf{148}, 095101 (2018).

\bibitem{manoj2018} J. Valiyakath and M. Gopalakrishnan, Polymerisation force of a rigid filament bundle: diffusive interaction leads to sublinear force-number scaling, Scientific Reports \textbf{8}, 2526 (2018).

\bibitem{Sadhu2016} R. K. Sadhu and S. Chatterjee, Actin filaments growing against a barrier with fluctuating shape, Phys. Rev. E {\bf 93}, 062414 (2016).

\bibitem{kirone} K. Tsekouras, D. Lacoste, K. Mallick and J. F. Joanny, Condensation of actin filaments pushing against a barrier, New J. Phys. {\bf 13}, 103032 (2011).

\bibitem{krawczyk2011} J. Krawczyk and J. Kierfeld, Stall force of polymerizing microtubules and filament bundles, Europhys. Lett. {\bf 93}, 28006 (2011).

\bibitem{ddas2014} D. Das, D. Das and R. Padinhateeri, Collective force generated by multiple biofilaments can exceed the sum of forces due to individual ones, New J. Phys. {\bf 16}, 063032 (2014).

\bibitem{hansda2014} D. K. Hansda, S. sen and R. Padinhateeri, Branching influences force-velocity curve and length fluctuations in actin networks, Phys. Rev. E {\bf 90}, 062718 (2014).

\bibitem{Sadhu2018} R. K. Sadhu and S. Chatterjee, Actin filaments growing against an elastic membrane: Effect of membrane tension, Phys. Rev. E {\bf 97}, 032408 (2018).

\bibitem{kolomeisky2015} X. Li and A. B. Kolomeisky, the role of multifilament structures and lateral interactions in dynamics of cytoskeleton proteins and assemblies, J. Phys. Chem. B \textbf{119}, 4653 (2015). 

\bibitem{peskin1993} C. S. Peskin, G. M. Odell and G. F. Oster, Cellular motions and thermal fluctuations: the Brownian ratchet, Biophys. J. {\bf 65}, 316 (1993).

\bibitem{mogilner1996} A. Mogilner and G. Oster, Cell motility driven by actin polymerization, Biophys. J. \textbf{71}, 3030 (1996).

\bibitem{AVolmer1998} A. Volmer, U. Seifert and R. Lipowsky, Critical behaviour of interacting surfaces with tension, Eur. Phys. J. {\bf 5}, 811 (1998).

\bibitem{lipowsky94} R. Lipowsky and S. Grotehans, Renormalization of hydration forces by collective protrusion modes, Biophys. Chem. {\bf 49} 27 (1994); R. Lipowsky and S. Grotehans, Hydration vs. Protrusion Forces Between Lipid Bilayers, Europhys. Lett. {\bf 23} 599 (1993).

\bibitem{baumgaertner2012} A. Baumgaertner, Crawling of a driven adherent membrane, J. Chem. Phys. {\bf 137}, 144906 (2012). 

\bibitem{pollard} T. D. Pollard, Rate constants for the reactions of ATP- and ADP-actin with the ends of actin filaments, J. Cell. Biol. \textbf{103}, 2747 (1986).

\bibitem{carlsson2008} A. E. Carlsson, Model of reduction of actin polymerization forces by ATP hydrolysis, Phys. Biol. \textbf{5}, 036002 (2008).

\bibitem{kpz} M. Kardar, G. Parisi and Y-C. Zhang, Dynamic scaling of growing interfaces, Phys. Rev. Lett. {\bf 56}, 889 (1986).
\end{thebibliography}
\end{document}